# Noise-Crypt: Image Encryption with Non-linear Noise, Hybrid Chaotic Maps, and Hashing

Laiba Asghar
*Department of Cyber Security, Pakistan Navy Engineering College*
*National University of Sciences and Technology*
Karachi, Pakistan
asgharlaiba609@gmail.com

Fawad Ahmed
*Department of Cyber Security, Pakistan Navy Engineering College*
*National University of Sciences and Technology*
Karachi, Pakistan
fawad@pnec.nust.edu.pk

Muhammad Shahbaz Khan
*School of computing, Engineering and the Built Environment*
*Edinburgh Napier University*
Edinburgh, United Kingdom
muhammadshahbaz.khan@napier.ac.uk

Arshad Arshad
*School of computing, Engineering and the Built Environment*
*Glasgow Caledonian University*
Glasgow, United Kingdom
arshad.arshad@gcu.ac.uk

Jawad Ahmad
*School of computing, Engineering and the Built Environment*
*Edinburgh Napier University*
Edinburgh, United Kingdom
j.ahmad@napier.ac.uk

*Abstract*—To secure the digital images over insecure transmission channels, a new image encryption algorithm Noise-Crypt is proposed in this paper. Noise-Crypt integrates non-linear random noise, hybrid chaotic maps, and SHA-256 hashing algorithm. The utilized hybrid chaotic maps are the logistic-tent and the logistic-sine-cosine map. The hybrid chaotic maps enhance the pseudorandom sequence generation and selection of substitution boxes, while the logistic-sine-cosine map induces non-linearity in the algorithm through random noise. This deliberate inclusion of noise contributes to increased resistance against cryptanalysis. The proposed scheme has been evaluated for several security parameters, such as differential attacks, entropy, correlation, etc. Extensive evaluation demonstrates the efficacy of the proposed scheme, with almost ideal values of entropy of 7.99 and correlation of -0.0040. Results of the security analysis validate the potency of the proposed scheme in achieving robust image encryption.

## I. Introduction

With the advancement of network technology, the volume of data exchanged over insecure networks has increased tremendously [1]. The ubiquity of such insecure data exchanges has given rise to an increase in cyber-attacks, which provides unauthorized access to digital images and discloses sensitive information to intruders [2]. These security concerns highlight the need to develop effective image encryption algorithms. Several cryptographic algorithms have been developed to ensure data confidentiality and integrity. Image data has specific characteristics, such as containing large amount of information and having high redundancy. Hence, the encryption schemes like AES that are widely used for data encryption are not suitable for images [3]. A secure encryption algorithm must have properties of diffusion and confusion. Confusion is a change in pixel location, whereas diffusion refers to a change in grey levels intensities [4].

Chaos theory plays critical role in image encryption due to its inherent characteristics, i.e., sensitivity to control parameters, nonlinearity, ergodicity, etc. Researcher prefer to integrate chaos with image encryption algorithms to increase security of the encryption schemes [5]. Chaotic maps introduce non-linearity into encryption schemes, making them difficult for intruders to crack [6, 7]. Traditionally chaotic maps are categorized as one dimensional and multidimensional maps. The one dimensional chaotic maps are simple and have limited key space [8]. Multi-dimensional maps have large chaotic range but have complex structures [9]. To minimize these disadvantages, researchers employ hybrid approaches combining various maps to improve 1D maps [10].

In chaotic encryption schemes, another important component is the substitution box (S-box) substitution. In traditional S-box substitution techniques, the mapping bijective, which means that each element of substitution box replaces the pixel of the original image at the same location as itself. Such technique is not effective for highly correlated data [11, 12]. So, the S-box substitution methods alone are not enough. A good encryption system should be able to scramble the image effectively and also should be difficult to crack [13]. In this regard, introducing random noise in an encryption algorithm enhances its unpredictability, complicating potential decryption attempts by adversaries. This added layer of randomness disrupts patterns, making it challenging to discern any underlying structure or sequence. As a result, the security robustness of the encryption scheme is significantly bolstered. Another important component of encryption techniques is usage of hash function. If a hash function is ideal in encryption schemes, it produces distinct outputs for different inputs and generates a fixed-size output for any given input [14, 15, 16].

This paper presents a new image encryption method that uses chaotic S-box substitution, adds random noise, and includes a hash function. This approach not only ensures efficient encryption of the palintext image but also enhances its resistance against unauthorized decryption attempts.

Main contributions of this paper are:

1. A novel image encryption scheme Noise-Crypt has been proposed. Noise-Crypt leverages random noise, which is generated through a hybrid chaotic map. This noise induces non-linearity and improves entropy of the proposed scheme making it more suitable for grayscale image encryption.

2. Integration of different chaotic maps to create improved hybrid maps. The logistic and tent maps have been combined to make a logistic-tent map, and the logistic, sine, and cosine maps have been combined to create a logistic-sine-cosine map. This increased the chaotic region of traditional maps.

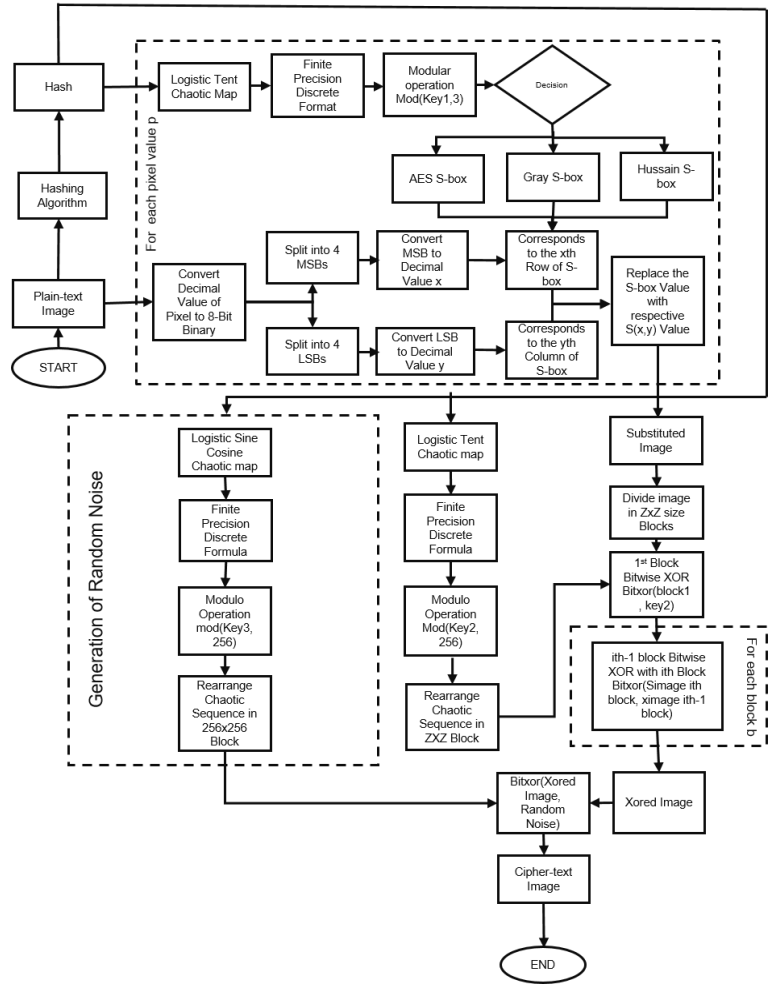

Fig. 1. The Proposed Encryption Scheme

3. The SHA-256 hash function has been employed to ensure that any slight change in the plaintext image affects the cipher images and keys, enhancing the overall security and integrity of our encryption approach.

## II. THE PRPOSED SCHEME – NOISE-CRYPT

Noise-Crypt uses properties of chaos, and hash to encrypt the images. A random noise is generated to introduce more randomness and improve the entropy. The scheme may be broken down into two algorithms. The algorithm 1 generates the keys for the S-box selection and Bit X-OR operations using two hybrid chaotic maps, whereas the algorithm 2 discusses the encryption process comprising of S-Box substitution, random noise, and Bit X-OR operation to generate the cipher image.

*A. Steps*

1. Read plain image of size MxN.

2. Generate hash sh_P of the plain image text using the sha256 algorithm and take first 11 characters of the hash then convert them to a decimal number d.

3. Set dd equals to d/10^14 to make sure the number lies in [0,1] range.

4. Set the initial value x0 to dd and initiate logistic tent map to generate key sequence K with the length of $(1, k_i)$, where i = M*N. Equation (1) represents logistic tent map mathematically:

$$\begin{cases} Xn+1 = \begin{pmatrix} rXn(1-Xn) + \\ (4-r)Xn/2 \end{pmatrix} mod1 & Xi < 0.5 \\ Xn+1 = \begin{pmatrix} rXn(1-Xn) + \\ (4-r)(1-Xn)/2 \end{pmatrix} mod1 & Xi \geq 0.5 \end{cases} \quad (1)$$

where parameter r ∈ (0, 4].

5. Multiply all the values of chaotic sequence with 10^14 for precision. Use round (K) to round the values to nearest integer.

6. Take mod (K,3) on all values to make sure no value is greater than 2 and we have 3 values to select for 3 S-boxes, use reshape operation reshape (K, M, N), where M and N are the dimensions of the image, to make sure the size of chaotic sequence is same as input image.

7. Calculate MSB and LSB of each pixel so that MSB corresponds to xth row and LSB corresponds to yth column of the S-box.

8. Generate AES, Hussain and Gray S-boxes for substitution and select the S-box randomly for substitution by chaotic sequence K to substitute the pixel value I(i,j) with selected S-box value Sn(i,j), where n is the number of S-box selected.

**ALGORITHM 1: KEY GENERATION ALGORITHM**

**Input:** a plain image of size Mimg x Nimg
**Output:** Encrypted image

1: I = readImage ("Image")
2: Sh_I = sha256(I)
3: $x_0$ = Sh_I
4: Set mu
5: **FOR** $j \leftarrow 1$ **to** Mimg × Nimg **do**
6:   **IF** $x(j) < 0.5$ **do**
    $x(j+1) \leftarrow mod\left(\begin{pmatrix} mu \times x(j) \times (1-x(j)) + \\ (4-mu) \times x(j)/2 \end{pmatrix}, 1\right)$
7:   **ELSE do**
    $x(j+1) \leftarrow mod\left(\begin{pmatrix} mu \times x(j) \times (1-x(j)) + \\ (4-mu) \times (1-x(j))/2 \end{pmatrix}, 1\right)$
8:   **END IF**
9: **END FOR**
10: $Key1 \leftarrow reshape\left(mod(round(X \times (10^{14})), 3), Mimg, Nimg\right)$
11: **FOR** $j \leftarrow 1$ **to** $Z^2$ **do**
12:   **IF** $x(j) < 0.5$ **do**
    $x(j+1) \leftarrow mod\left(\begin{pmatrix} mu \times x(j) \times (1-x(j)) + \\ (4-mu) \times x(j)/2 \end{pmatrix}, 1\right)$
13:   **ELSE do**
    $x(j+1) \leftarrow mod\left(\begin{pmatrix} mu \times x(j) \times (1-x(j)) + \\ (4-mu) \times (1-x(j))/2 \end{pmatrix}, 1\right)$
14:   **END IF**
15: **END FOR**
16: $Key2 \leftarrow reshape(mod(round(X \times (10^{14})), 256), Z, Z)$
17: **FOR** $j \leftarrow 1$ **to** Mimg × Nimg **do**
    $x(j+1) \leftarrow \cos\left(pi\begin{pmatrix} 4 \times mu \times x(j) \times (1-x(j)) + \\ (1-mu) \times \sin(pi \times x(j)) - 0.5 \end{pmatrix}\right)$
18: **END FOR**
19: $Key3 \leftarrow reshape(mod(round(X \times (10^{14})), 256), Mimg, Nimg)$

9. Divide the $S_{image}$ into fix ZxZ blocks.

10. Initiate the logistic tent map again to generate another key sequence K2 of length (1, Z) where Z is the length of the block, the initial value x0 is equal to dd. Equation (2) represents logistic tent map mathematically:

$$\begin{cases} Xn+1 = \begin{pmatrix} rXn(1-Xn) + \\ (4-r)Xn/2 \end{pmatrix} mod\ 1 & Xi < 0.5 \\ Xn+1 = \begin{pmatrix} rXn(1-Xn) + \\ (4-r)(1-Xn)/2 \end{pmatrix} mod\ 1 & Xi \geq 0.5 \end{cases} \quad (2)$$

where parameter r ∈ (0, 4].

11. Multiply all the values of chaotic sequence with 10^14 for precision. Use round (K2) to round the values to nearest integer.

12. Took mod (K,256) on all values to make sure no value is greater than 255 so they are same as the pixel values. Use reshape operation reshape (K, Z, Z).

13. The first block b1 of $S_{image}$ (1:Z, 1:Z) is Bit X-ORed with the K2. $bitxor(block1, K2)$.

14. The next block b2 is Bit X-ORed with the output of previous block The same procedure is followed by all b(m*n/z) blocks. $bitxor\ (Subimage\ ith\ block, Xored\ image\ ith - 1\ block)$

15. This step gives us the Xored image $X_{image}$.

16. Initiate the logistic sine cosine map to generate random noise RN of length (1, MxN), the initial value x0 is equal to dd. Logistic sine cosine map can be represented mathematically as (3)

$$Xn+1 = cos\left(pi\begin{pmatrix} 4rXn(1-Xn) + \\ (1-r)\sin(piXn) - 0.5 \end{pmatrix}\right) \quad (3)$$

Where parameter r ∈ [0, 1]

17. Multiply all the values of random noise with 10^14 for precision. Use round (RN) to round the values to nearest integer.

18. Take mod (RN,256) on all values to make sure no value is greater than 2155 so that all values are same as the pixel values to perform Bit X-OR operation.

19. Use reshape operation reshape (RN, M, N) to reshape random noise into 256x256 matrix.

20. The random noise RN and xored image Ximage are bitwise xored to get ciphertext image.

## III. RESULTS AND ANALYSIS

We have compared the proposed algorithm to approaches from relevant literature. Several security analysis metrics, such as the cipher image histogram, entropy, contrast, correlation, energy, homogeneity, NPCR, UACI are offered in this part to assess the effectiveness of the proposed Chaotic encryption approach. All tests are run on plaintext picture and their associated cipher images.

### A. Encrypted Image Analysis

The most common security analysis is the comparison of the histograms of both the original and encrypted images. A histogram shows the distribution of pixel brightness in an image, representing how many pixels have each intensity level [17]. This gives insights into the image's brightness and contrast. When we encrypt a picture, we change its pixel values to make the image secure and hard to read. By looking at the histogram, we can see how the encryption changes the distribution of pixel intensities.

**ALGORITHM 2: ENCRYPTION PROCESS**

**Input:** a plain image of size Mimg × Nimg
**Output:** Encrypted Image

1: **FOR** $i \leftarrow 1$ **to** Mimg **and FOR** $j \leftarrow 1$ **to** Nimg **do**
2:   %Calculate MSB and LSB to locate indices of S boxes
3:   **IF** $Key1(i,j) \leftarrow 0$ **do**
4:     $Subimage(i,j) \leftarrow Substitute\ (I(i,j), S1(i,j))$
5:   **ELSEIF** $Key1(i,j) \leftarrow 1$ **do**
6:     $Subimage(i,j) \leftarrow Substitute\ (I(i,j), S2(i,j))$
7:   **ELSE do**
8:     $Subimage(i,j) \leftarrow Substitute\ (I(i,j), S3(i,j))$
9:   **END IF**
10: **END FOR**
11: $block\ 1 \leftarrow bitxor(Subimage(1:Z, 1:Z), Key2)$
12: $XoredImage(1:Z, 1:Z) \leftarrow block1$
13: **FOR** $i \leftarrow Z+1$ **to** Mimg **and FOR** $j \leftarrow Z+1$ **to** Nimg **do**
14:   $XoredImage(i,j) \leftarrow bitxor\begin{pmatrix} Subimage(i,j), \\ XoredImage(i-Z, j-Z) \end{pmatrix}$
15: **END FOR**
16: CipherImage ← bitxor (XoredImage(i,j), Key3)

Fig. 2. Results of Noise-Crypt; (a-b) Cameraman image with its histogra, (c-d) Encrypted image with its histogram.

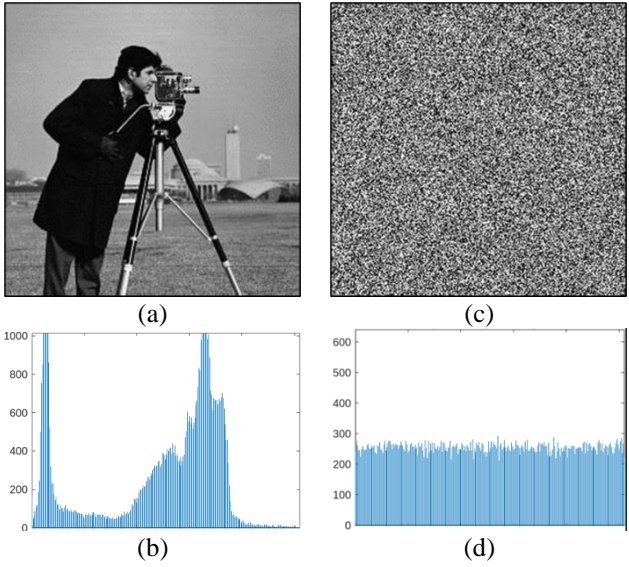

(a)     (c)

(b)     (d)

TABLE I. SECURITY RESULTS OF PROPOSED ENCRYPTION SCHEME

| Security Parameter | Plain Image | Cipher Image |
|---|---|---|
| Contrast | 0.3783 | 10.5505 |
| Entropy | 7.1028 | 7.9977 |
| Correlation | 0.9501 | -0.0040 |
| Homogeneity | 0.8990 | 0.3893 |
| Energy | 0.1818 | 0.0156 |

The encrypted images display histograms that are evenly distributed, showing that the proposed algorithm effectively hides the content of the original image and stands strong against attacks based on histogram analysis. We can measure this using the chi-square ($\chi^2$) test. A lower chi-square value indicates a more uniform distribution. The chi-square is mathematically defined in equation (4):

$$chi\ square = \sum_{i=0}^{255} \frac{(f_i \times \varepsilon)^2}{\varepsilon} \quad (4)$$

where $\varepsilon = \frac{M \times N}{256}$

Where $f_i$ represents cipher images histogram values at index i, M and N are height and width of the image. Below are the histograms of plain and cipher images.

### B. Contrast

The intensity difference between a pixel and its neighbor is measured via contrast analysis throughout the entire picture. High contrast value indicates secure algorithm. When we encrypt an image, the randomness in the image increases and in response the contrast levels reach a very high value [19]. A greater contrast implies more randomly placed pixels in the image. Mathematically contrast is represented in equation (5):

$$Contrast = \sum_{i,j} |i-j|^2 p(i,j) \quad (5)$$

Where i and j are intensity levels.

### C. Entropy

Entropy measures the amount of uncertainty in a dataset. It refers to the degree of unpredictability or fluctuation in pixel values of images. A higher entropy number implies greater uncertainty or unpredictability, whereas a lower entropy value indicates greater predictability and less randomness.

Mathematically entropy is represented as (6):

$$H = \sum_{i=0}^{N-1} p(i) \times \log_2 \frac{1}{p(i)} \quad (6)$$

where $N$ = gray level count, and $p(i)$ = pixel likelihood having value $i$.

Information entropy measures the randomness or unpredictability of information. In a chaotic encryption scheme, the closer the entropy value is to 8, the more chaotic it is [18]. The ideal entropy can be determined using equation (7):

$$H_{ideal} = \sum_{i=0}^{255} \frac{1}{256} \times \log_2 \frac{1}{256} = 8 \quad (7)$$

A low entropy value is advantageous for attackers and the cipher images with low entropy are sensitive to cryptanalysis and brute force attack

### D. Correlation

In image encryption, "correlation" means how pixel values in an image relate to each other. It's best if nearby pixels in the encrypted image don't have predictable or similar values. This makes the encrypted image look random. If two encrypted images have a correlation value close to 0, they don't share obvious patterns and are considered very different from each other.

Mathematically correlation coefficient cab be represented as follows:

$$CorrC = \frac{\sum_{i=1}^{M}\sum_{j=1}^{N}\binom{(P(i,j)-E(P))}{(C(i,j)-E(C))}}{\sqrt{\sum_{i=1}^{M}\sum_{j=1}^{N}(P(i,j)-E(P))^2 \sum_{i=1}^{H}\sum_{j=1}^{N}(C(i,j)-E(C))^2}} \quad (8)$$

Where $P(i,j)$ and $C(i,j)$ are the pixel values at specific location. $E(P)$ and $E(C)$ are the expected values.

### E. Homogenity

Homogeneity is a derived statistic of the GLCM matrix. Gray Level Co-occurrence Matrix (GLCM) is a matrix that describes spatial relationship between intensity levels of pixels at a given offset in an image. Homogeneity describes the similarity between adjacent pixel values. Encryption algorithms with low homogeneity values are considered secure because the goal of encryption is to have high randomness and high homogeneity indicates more uniformity and less randomness.

Mathematically homogeneity can be written as (9):

$$Homogenity = \sum_{i,j} \frac{P(i,j)}{1+|i-j|} \quad (9)$$

Where P(i,j) is the pixel value at ith row and jth column.

### F. Energy

Energy, a statistic derived from the GLCM, is also called uniformity. It represents the sum of the squared elements in the GLCM. [20]. Algorithms with low energy values are considered secure.

Mathematically energy can be represented as (10):

$$Energy = \sum_{i,j} P(i,j)^2 \qquad (10)$$

Where i and j are two adjacent intensity levels.

*G. NPCR and UACI*

NCPR calculates the difference between two ciphertext images. One is the ciphertext of the non-tampered plaintext image, the other is the tampered plaintext image. We can use this measure to see how well an encryption method stands up to differential attacks.

Mathematically NCPR is given as:

$$NCPR = \sum_{i=1}^{M} \sum_{j=1}^{N} D(i,j) \qquad (11)$$

Where $D = \begin{cases} 0 \; if \; C_2(i,j) = C_1(i,j) \\ 1 \; if \; C_2(i,j) \neq C_1(i,j) \end{cases}$

M represents the height and N is the width of the image. C1 is the encrypted version of the original image, and C2 is the encrypted version of that image when just 1 bit is altered. A higher NPCR means the encryption method is more secure, as it shows high sensitivity to minor changes in pixel values. For our method, the NPCR is 99.58%. Another measure, Unified Average Change Intensity (UACI), is used in image encryption to find out the average intensity difference between two encrypted images that come from the same original image with a slight one-bit difference.

$$UACI = \frac{1}{H \times W} \left[ \frac{\sum_{i=1}^{M} \sum_{j=1}^{N} |C_1(i,j) - C_2(i,j)|}{2^b - 1} \right] \times 100\% \qquad (10)$$

In this context, M stands for the height and N for the width of the image. C1 represents the encrypted version of the original image, while C2 is the encrypted version when just a single bit of the original image is altered. Additionally, b represents the number of bits in each image pixel.

## IV. CONCLUSION

This paper introduced and evaluated a novel encryption scheme that combines hybrid chaotic maps, random noise, and the SHA-256 hashing algorithm. The chaotic map-generated random noise increased the non-linearity of the algorithm. To further improve security, the key parameters for the chaotic maps were drawn directly from the original image, ensuring that keys were both more secure and dependent on the plaintext. Our security evaluations revealed that our image encryption technique, which leverages S-Box selection and chaotic map-driven random noise, outperformed others, particularly concerning statistical security metrics.